\documentclass[preprint,twocolumn]{aastex631}

\usepackage[T1]{fontenc}
 \usepackage{pslatex}
\usepackage{amsmath}
\usepackage{amsfonts}
\usepackage{color,ulem}
\usepackage{url}
\usepackage{graphicx}
\usepackage{subfigure}
\usepackage{amssymb}
\usepackage{soul}
\usepackage{booktabs}
\usepackage{array}
\usepackage[utf8]{inputenc}
\inputencoding{latin1}
\inputencoding{utf8}

{\newif\ifnotend
\notendtrue
\def\veclist{ABCDEFGHIJKLMNOPQRSTUVWXYZabcdefghijklmnopqrstuvwxyz.}
\def\top#1#2.{#1}
\def\tail#1#2.{#2.}
\loop\expandafter\xdef\csname v\expandafter\top\veclist\endcsname%
{{\noexpand\bf\expandafter\top\veclist}}
\edef\veclist{\expandafter\tail\veclist}
\if\veclist.\notendfalse\fi\ifnotend\repeat}
\def\d{{\rm d}}
\def\e{{\rm e}}
\def\i{{\rm i}}

\def\E{{\cal E}}

\mathchardef\mhyphen="2D

\begin{document}

\title{Ultrafast Variability in AGN Jets: Intermittency and Lighthouse Effect}
\shorttitle{Ultrafast Variability in AGN Jets}

\correspondingauthor{Emanuele Sobacchi}
\email{emanuele.sobacchi@mail.huji.ac.il}

\shortauthors{Sobacchi, Piran, and Comisso}

\author{Emanuele Sobacchi}
\affiliation{Racah Institute for Physics, The Hebrew University, Jerusalem 91904, Israel}

\author{Tsvi Piran}
\affiliation{Racah Institute for Physics, The Hebrew University, Jerusalem 91904, Israel}

\author{Luca Comisso}
\affiliation{Department of Astronomy and Columbia Astrophysics Laboratory, Columbia University, New York, NY 10027, USA}

\def\p{\partial}
\def\E{\textbf{E}}
\def\B{\textbf{B}}
\def\v{\textbf{v}}
\def\j{\textbf{j}}
\def\s{\textbf{s}}
\def\e{\textbf{e}}

\newcommand{\di}{\mathrm{d}}
\newcommand{\bfx}{\mathbf{x}}
\newcommand{\bfe}{\mathbf{e}}
\newcommand{\vlos}{\mathrm{v}_{\rm los}}
\newcommand{\Tspin}{T_{\rm s}}
\newcommand{\Tb}{T_{\rm b}}
\newcommand{\degree}{\ensuremath{^\circ}}
\newcommand{\Th}{T_{\rm h}}
\newcommand{\Tc}{T_{\rm c}}
\newcommand{\bfr}{\mathbf{r}}
\newcommand{\bfv}{\mathbf{v}}
\newcommand{\bfu}{\mathbf{u}}
\newcommand{\pc}{\,{\rm pc}}
\newcommand{\kpc}{\,{\rm kpc}}
\newcommand{\Myr}{\,{\rm Myr}}
\newcommand{\Gyr}{\,{\rm Gyr}}
\newcommand{\kms}{\,{\rm km\, s^{-1}}}
\newcommand{\de}[2]{\frac{\partial #1}{\partial {#2}}}
\newcommand{\cs}{c_{\rm s}}
\newcommand{\rb}{r_{\rm b}}
\newcommand{\rqu}{r_{\rm q}}
\newcommand{\bfOmega}{\pmb{\Omega}}
\newcommand{\bfOmegap}{\pmb{\Omega}_{\rm p}}
\newcommand{\bfXi}{\boldsymbol{\Xi}}

 \newcommand\es[1]{{\color{red} #1}}
 \newcommand\tp[1]{{\color{blue} #1}} 
  \newcommand {\dtp}[1] {\textcolor{red}{\sout{TP #1}}}
\newcommand{\rtp}[2]  {{\textcolor{red}{\sout{#1}}}{\textcolor{blue}{ #2}}}

\begin{abstract}
Gamma-ray flares from Active Galactic Nuclei (AGN) show substantial variability on ultrafast timescales (i.e.\ shorter than the light crossing time of the AGN's supermassive black hole). We propose that ultrafast variability is a byproduct of the turbulent dissipation of the jet Poynting flux. Due to the intermittency of the turbulent cascade, the dissipation is concentrated in a set of reconnecting current sheets. Electrons energised by reconnection have a strong pitch angle anisotropy, i.e.\ their velocity is nearly aligned with the guide magnetic field. Then each current sheet produces a narrow radiation beam, which dominates the emission from the whole jet when it is directed towards the observer. The ultrafast variability is set by the light crossing time of a single current sheet, which is much shorter than the light crossing time of the whole emission region. The predictions of our model are: (i) The bolometric luminosity of ultrafast AGN flares is dominated by the inverse Compton (IC) emission, as the lower energy synchrotron emission is suppressed due to the pitch angle anisotropy. (ii) If the observed luminosity includes a non-flaring component, the variations of the synchrotron luminosity have a small amplitude. (iii) The synchrotron and IC emission are less variable at lower frequencies, as the cooling time of the radiating particles exceeds the light crossing time of the current sheet. Simultaneous multiwavelength observations of ultrafast AGN flares can test these predictions.
\end{abstract}



\section{Introduction}

Relativistic jets from Active Galactic Nuclei (AGN) are observed across the entire electromagnetic spectrum \citep[for a review see][]{Blandford+2019}. The observed non-thermal emission exhibits substantial variability on a wide range of timescales, from several years down to a few minutes. The ultrafast variability (i.e.\ faster than the light crossing time of the supermassive black hole Schwarzschild radius) is particularly interesting as it cannot be imprinted by the engine, and therefore can enlighten the physics of non-thermal particle acceleration in the jet.

Ultrafast variability of the gamma-ray emission has been reported in several AGN flares \citep[e.g.][]{Albert+2007, Aharonian+2007, Aleksic+2011, Aleksic+2014, Ackermann+2016}. The detection of an ultrafast TeV flare from the Flat Spectrum Radio Quasar PKS 1222+21 places the flaring site at a distance $\gtrsim 0.1{\rm\; pc}$, as required to prevent absorption of the gamma-rays by the UV photons from the Broad Line Region \citep[][]{Aleksic+2011}.

Most theoretical models of ultrafast gamma-ray flares belong to three main categories \citep[for a review see][]{AharonianBarkov2017}:
\begin{enumerate}
\item Models where ultrafast gamma-ray flares are produced in magnetospheric gaps/current sheets \citep[e.g.][]{Neronov2007, Ghisellini+2009, Levinson2011, Crinquand+2020, Crinquand+2021, Hakobyan+2023}. These models cannot explain the origin of the ultrafast TeV flare from PKS 1222+21, as the emission radius is necessarily large, while they remain viable for other sources.
\item Models where ultrafast gamma-ray flares are produced due to the interaction of the jet with an external cloud or a star \citep[e.g.][]{Araudo+2010, Barkov+2012}. However, this scenario requires a very large jet power \citep[][]{AharonianBarkov2017}.
\item Models attributing the ultrafast variability to the relativistic random motion of ``blobs'' in the proper frame of the jet. These ``jets-in-a-jet'' models were developed for Gamma-Ray Bursts (GRBs) \citep[e.g.][]{Lyutikov2006, Lazar+2009, NarayanKumar2009} and later applied to AGN \citep[e.g.][]{Giannios2009, Nalewajko2011, NarayanPiran2012, Giannios2013}. 
\end{enumerate}

\citet[][]{Giannios2009} \citep[see also][]{Giannios2013} pointed out that the jets-in-a-jet scenario can be realised when the jet energy is dissipated via magnetic reconnection (however, \citet{NarayanPiran2012} find strong constraints on the feasibility of this model). In this model the reconnection layer is fragmented into a chain of blobs -- the so-called plasmoids. If the magnetic field across the current sheet is nearly anti-parallel, the plasmoids can be accelerated to ultrarelativistic velocities 
\citep{Lyubarsky2005, Giannios2009}. An important prediction of such reconnection-driven jets-in-a-jet scenario is that ultrafast gamma-ray flares should be accompanied by bright ultrafast X-ray flares \citep[][]{Giannios2009, Petropoulou+2016, Christie+2019}. These two components originate from the inverse Compton (IC) and synchrotron emission of the radiating electrons. The synchrotron losses are expected to be large as the plasma is strongly magnetised.

\citet[][]{Thompson2006} proposed a different scenario to produce the fast variability of GRBs. In this model the jet energy is dissipated via a turbulent Alfv\'{e}nic cascade. The intermittency of the cascade is neglected, and consequently the dissipation rate in the emission region is uniform. \citet[][]{Thompson2006} argues that particles are energised via Landau damping. The energised particles have a strong pitch angle anisotropy, i.e.\ their velocity is nearly aligned with the direction of the local magnetic field. Since the emitted radiation is beamed along the field, the observer sees a small part of the emission region, where the field is nearly aligned with the line of sight. This effect produces fast variability.

In this paper we revisit the scenario of \citet[][]{Thompson2006}, and apply it to AGN jets. Our work is motivated by recent kinetic simulations of relativistic plasma turbulence \citep{ComissoSironi2018, ComissoSironi2019}. These simulations show that the turbulent cascade is intermittent, and the dissipation is concentrated in a set of large scale reconnecting current sheets. The reconnection magnetic field is not anti-parallel, and a relatively large guide field is present. In this case, the bulk motion of the plasmoids remains mildly relativistic, and the velocity of the electrons energised by reconnection is nearly aligned with the local magnetic field.\footnote{Instead, in the reconnection-driven jets-in-a-jet model, the magnetic field across the current sheet is nearly anti-parallel. Then the velocity of the plasmoids is ultrarelativistic, and the pitch angle distribution is isotropic.} Similarly to the original idea of \citet[][]{Thompson2006}, in our scenario the ultrafast variability arises due to the pitch angle anisotropy of the radiating particles. However, we show that the intermittency of the turbulent cascade is essential for the variability.

A natural prediction of our scenario is that ultrafast gamma-ray flares should have a faint counterpart at lower frequencies, as the synchrotron emission is suppressed when the velocity of the radiating particles is nearly aligned with the local magnetic field. Simultaneous multiwavelength observations of ultrafast gamma-ray flares can test this prediction, and discriminate between our scenario and the reconnection-driven jets-in-a-jet model considered in earlier work \citep[][]{Giannios2009, Petropoulou+2016, Christie+2019}, as the latter predicts bright ultrafast X-ray flares.

The paper is organised as follows. In Section \ref{sec:model} we present our scenario for the ultrafast variability of AGN flares. In Section \ref{sec:LC} we discuss the properties of the light curve. In Section \ref{sec:comparison} we compare our scenario with previous work. We summarise our results and conclude in Section \ref{sec:conclusions}.

\section{Origin of the variability}
\label{sec:model}

According to a widely accepted paradigm, relativistic jets extract the rotational energy of supermassive black holes via electromagnetic stresses \citep[][]{BlandfordZnajek1977, Tchekhovskoy2011}. In this process most of the enegy is carried in the form of Poynting flux. Due to the unsteady activity of the engine, the jet power may vary on a timescale
\begin{equation}
\label{eq:Teng}
\Delta T_{\rm eng}\gtrsim\frac{R_{\rm g}}{c}\;,
\end{equation}
where $R_{\rm g}/c=2GM/c^3=10^4 (M/10^9 M_\odot ) {\rm\; s}$ is the light crossing time of the engine Schwarzschild radius. Of course, the jet power cannot vary on timescales shorter than $R_{\rm g}/c$.

The jet Poynting flux can be dissipated as a result of magneto-hydrodynamical (MHD) instabilities, such as the kink instability and the Kelvin-Helmoltz instability. Consequently, a population of non-thermal particles is accelerated \citep[][]{Alves+2018, Davelaar+2020, Sironi+2021b}. If the cooling time of these particles is shorter than the expansion time of the jet (fast cooling regime), a substantial fraction of the jet power is converted into synchrotron and IC radiation.

A radiative flare can be produced by a shell that carries a large Poynting flux. We assume that the Poynting flux is dissipated while the shell moves between the distances $R_{\rm diss}$ and $R_{\rm diss}+\Delta R_{\rm diss}$ from the engine, where $\Delta R_{\rm diss}\sim R_{\rm diss}$. We assume that the jet has a nearly cylindrical shape and neglect the angular spreading.\footnote{The jet opening angle depends the radial profile of the external pressure that collimates the jet. If the pressure is independent of the distance from the engine, the jet has a nearly cylindrical shape \citep[][]{Lyubarsky2008}.} The photons emitted when the head of the shell reaches the distance $R_{\rm diss}$ will be received first, and the photons emitted when the tail of the shell reaches the distance $R_{\rm diss}+\Delta R_{\rm diss}$ will be received last. The observed duration of the flare, $T$, can be estimated from the difference of the arrival times. If the shell has a width $c\Delta T_{\rm eng}$ in the observer's frame, one finds \citep[][]{SariPiran1997}
\begin{equation}
\label{eq:Tflare}
T\sim \Delta T_{\rm eng} + \frac{R_{\rm diss}}{2\Gamma^2c} \;,
\end{equation}
where $\Gamma$ is the bulk Lorentz factor of the shell (it is assumed that $\Gamma\gg 1$). From Eqs.\ \eqref{eq:Teng} and \eqref{eq:Tflare} one sees that the duration of the flare cannot be shorter than the light crossing time of the engine, i.e.\ $T\gtrsim R_{\rm g}/c$.

\begin{figure}{\vspace{2mm}} 
\centering
\includegraphics[width=0.46\textwidth]{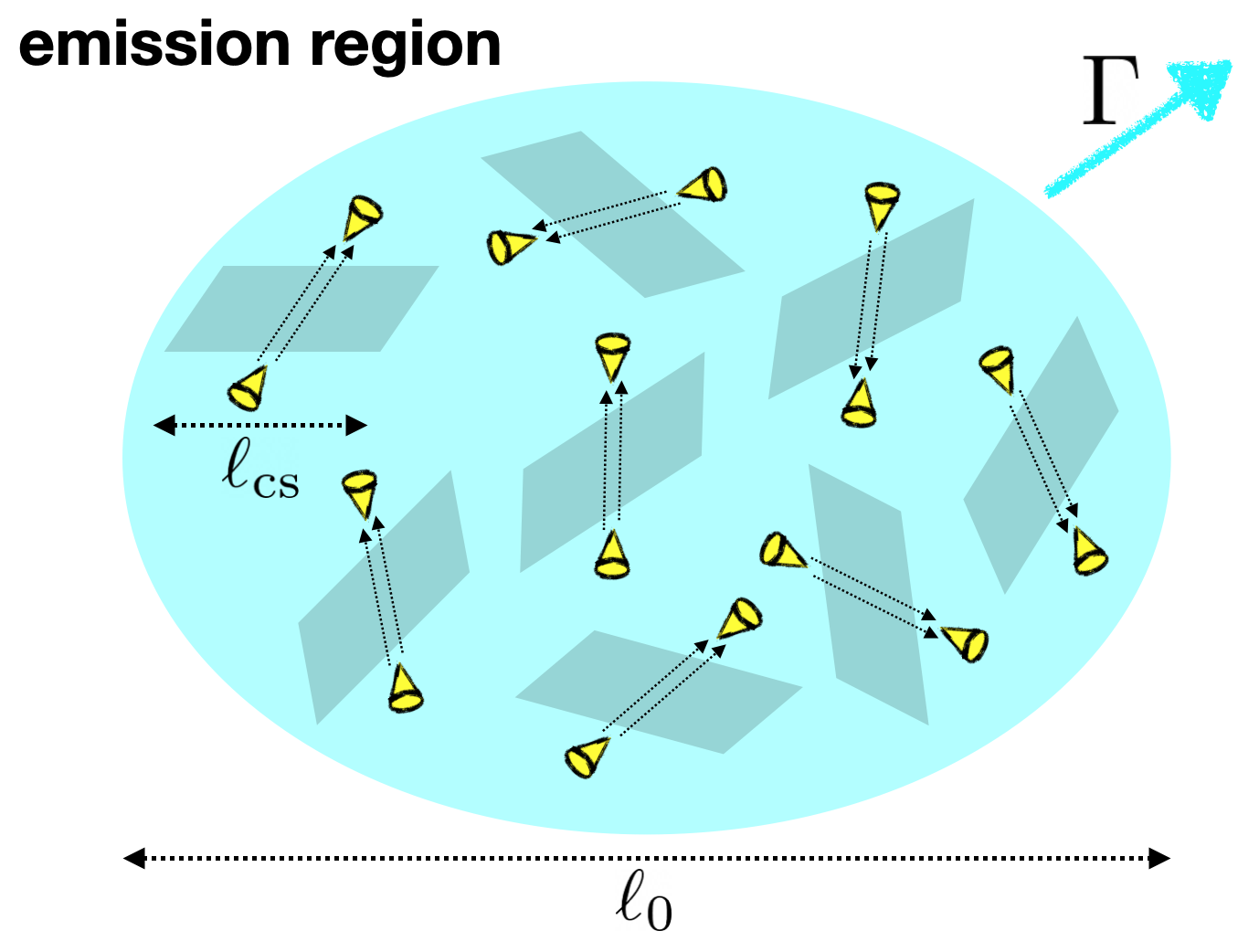}
\includegraphics[width=0.46\textwidth]{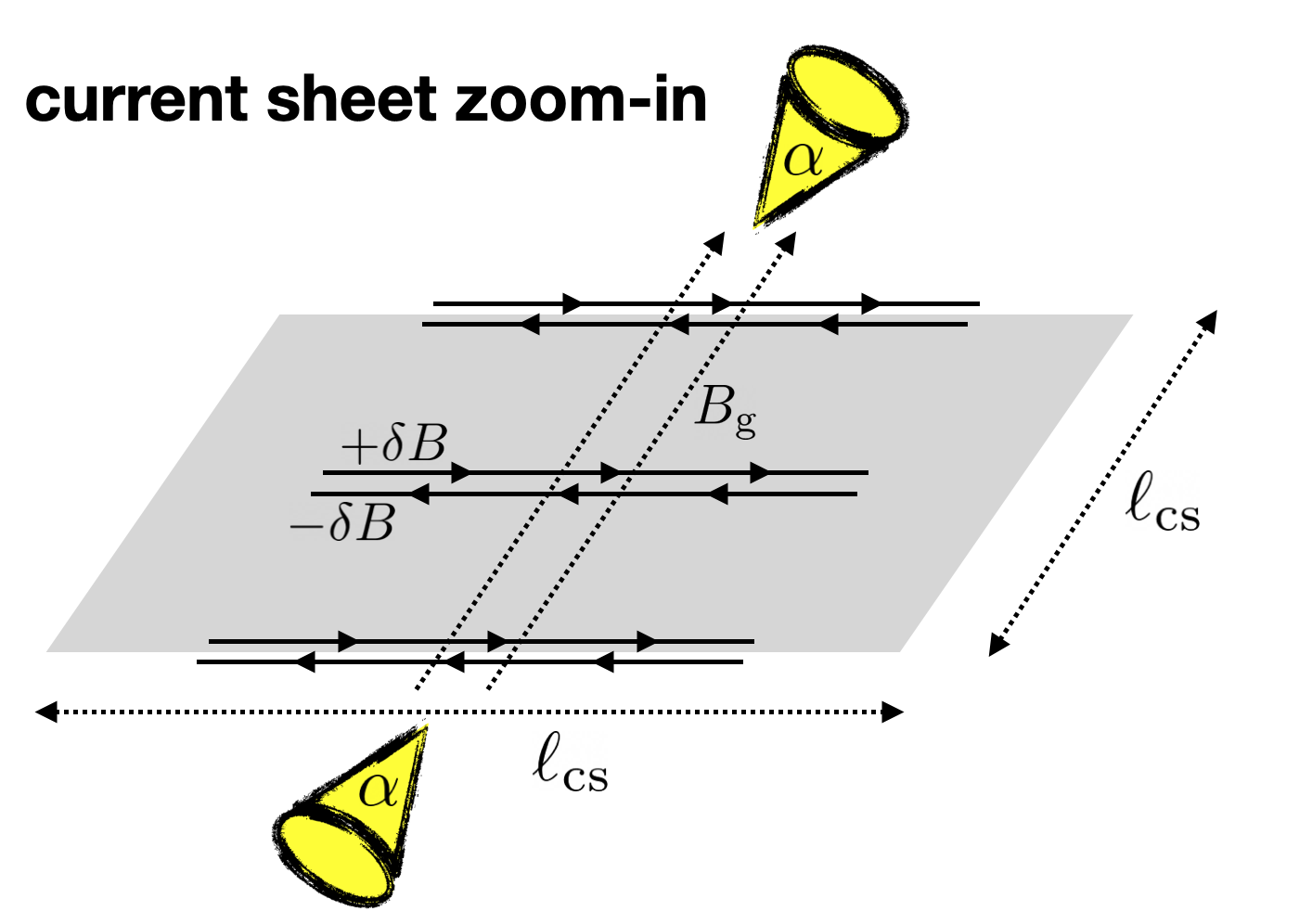}
\vspace{-2mm}
\caption{Cartoon of our scenario for the ultrafast variability of AGN flares. The jet's energy is dissipated via a turbulent cascade. Due to the intermittency of the cascade, the dissipation is concentrated in a set of reconnecting current sheets with a typical size $\ell_{\rm cs}$. Since the guide magnetic field is comparable to the anti-parallel component (i.e.\ $B_{\rm g}\sim\delta B$), electrons energised by reconnection have a strong pitch angle anisotropy, i.e.\ their velocity is nearly aligned with the guide field. Then each current sheet produces a narrow radiation beam with opening angle $\alpha\ll 1$ and duration $\tau_{\rm cs}\sim\ell_{\rm cs}/c$, which is much shorter than the total duration of the flare, $\tau\sim\ell_0/c$ (the duration is calculated in the proper frame of the emission region). The gamma-ray light curve is highly variable as a single beam dominates over the whole emission region when it intercepts the line of sight. Our model predicts that ultrafast gamma-ray flares have a faint counterpart at lower frequencies (optical, X-rays), as the synchrotron emission is suppressed due to the pitch angle anisotropy of the radiating particles. 
}
\label{fig:scenario}
\end{figure}

Below we show that the variability timescale of the light curve, $\delta T$, can be significantly shorter than the total duration of the flare, $T$. We outline the following scenario, which is illustrated in Figure \ref{fig:scenario}.  Since astrophysical jets have huge magnetic Reynolds numbers, it is conceivable that the jet Poynting flux is dissipated via a turbulent cascade. The cascade becomes increasingly intermittent at smaller scales, and the dissipation is concentrated in a set of reconnecting current sheets that fill a small fraction of the shell volume \citep[][]{Biskamp2003, ComissoSironi2018, ComissoSironi2019}. In the fast cooling regime, electrons energised by reconnection radiate their energy before moving far away from the current sheet \citep[][]{ComissoSironi2021, NattilaBeloborodov2021, SobacchiNattila2021}. Since the electrons have a strong pitch angle anisotropy, and move nearly along the direction of the guide magnetic field \citep[][]{ComissoSironi2019, ComissoSironi2021, Comisso+2020, SobacchiNattila2021}, the radiation of the current sheet is focused into a narrow beam. Then each current sheet behaves similarly to a lighthouse, and can be observed only if the field is directed nearly along the line of sight. This effect increases the apparent luminosity of the current sheets, and exacerbates the variability of the light curve. The variability timescale is equal to the light crossing time of the current sheets, while the duration of the flare is equal to the light crossing time of the entire shell, which can be significantly longer.

\subsection{Intermittency of the turbulent cascade}

Our goal is estimating the ratio of the variability timescale and the duration of the flare, $\delta T/T$, and the number of pulses of the light curve, $N_{\rm p}$. Since these quantities are relativistic invariant, we work in the proper frame of the shell. For simplicity we assume that\footnote{Our assumption could be realised if the engine launches two shells with a time delay $\Delta T_{\rm eng}\sim R_{\rm g}/c$. If the diffenence of the shells Lorentz factors is $\Delta\Gamma\sim\Gamma$, the shells collide at the distance $R_{\rm diss}\sim 2\Gamma^2c\Delta T_{\rm eng}$.} $R_{\rm diss}/2\Gamma^2c \sim \Delta T_{\rm eng}\sim R_{\rm g}/c$. Then Eq.\ \eqref{eq:Tflare} gives $T\sim\Delta T_{\rm eng}\sim R_{\rm g}/c$.

The volume of the shell calculated in the proper frame is $\mathcal{V}_0\sim \Gamma c\Delta T_{\rm eng}\ell_0^2$, where $\Gamma c\Delta T_{\rm eng}$ is the longitudinal length and $\ell_0^2$ is the transverse section. The magnetic energy of the shell is dissipated on the timescale $\tau\sim \Gamma T\sim \Gamma\Delta T_{\rm eng}$ (where $\tau$ is calculated in the proper frame of the shell). Due to the intermittency of the turbulent cascade, the dissipation occurs in a set of reconnecting current sheets. Numerical simulations show that the reconnecting component of the magnetic field is comparable to the guide field, which ensures a reconnection rate $\beta_{\rm rec}\sim 0.1$ \citep[][]{ComissoSironi2018, ComissoSironi2019}. The reconnection rate is associated with the inflow velocity of the plasma into the current sheets, which is $v_{\rm in}\sim \beta_{\rm rec}v_{\rm A}$, where $v_{\rm A}$ is the upstream Alfv\'{e}n velocity (in the relativistic regime, one has $v_{\rm A}\sim c$).

For simplicity we assume that the current sheets where the energy is dissipated are identical, each having surface area $\ell_{\rm cs}^2$. Let $N_{\rm cs}$ be the total number of these sheets, and $\tau_{\rm cs}$ the survival time of a sheet before being destroyed by the turbulent motions of the plasma. The condition that the current sheets process the entire volume of the shell gives $N_{\rm cs}v_{\rm in}\tau_{\rm cs}\ell_{\rm cs}^2\sim\mathcal{V}_0$. Taking into account that $v_{\rm in}\sim\beta_{\rm rec}v_{\rm A}\sim\beta_{\rm rec}c$, the total number of current sheets can be estimated as
\begin{equation}
\label{eq:Ns}
N_{\rm cs}\sim \beta_{\rm rec}^{-1}\left(\frac{\ell_0}{\ell_{\rm cs}}\right)^2\left(\frac{\tau}{\tau_{\rm cs}}\right) \;.
\end{equation}
Even if the current sheets are essentially two-dimensional structures that fill a small fraction of the shell volume, our model does not have an efficiency problem as the current sheets dissipate a large fraction of the magnetic energy of the shell.\footnote{This is essentially the same argument that can be made for fractal models of hydrodynamical turbulence intermittency, when the cascade accumulates on structures of fractal dimension $D<3$ \citep[][]{Frisch1995}.}

Models of Alfv\'{e}nic turbulence suggest that current sheets of scale $\ell_{\rm cs}$ are destroyed on the timescale $\tau_{\rm cs}\sim \ell_{\rm cs}/v_{\rm A}\sim\ell_{\rm cs}/c$ \citep[e.g.][]{Boldyrev2006}. We consider the regime where the cooling time of the energised particles, $\tau_{\rm cool}$, is shorter than the light crossing time of the sheets, $\tau_{\rm cs}$, and postpone the discussion of a less efficient cooling to Section \ref{sec:cooling}. When $\tau_{\rm cool}\lesssim\tau_{\rm cs}$, each current sheet produces a radiation pulse of duration $\tau_{\rm cs}$. Since the ratio of the duration of the pulse and the duration of the flare is relativistic invariant, one finds
\begin{equation}
\label{eq:tvar}
\frac{\delta T}{T}\sim\frac{\tau_{\rm cs}}{\tau}\sim \eta \frac{\ell_{\rm cs}}{\ell_0} \;,
\end{equation}
where $\eta=\ell_0/c\tau\sim\ell_0/\Gamma c\Delta T_{\rm eng}$ is the aspect ratio of the shell in the proper frame (it is assumed that $\eta\sim 1$).

Our Eq.\ \eqref{eq:tvar} is similar to Eq.\ (9) of \citet[][]{NarayanPiran2012}, but in their model the variability timescale is longer by a factor $\beta_{\rm rec}^{-1}$. The reason is that we adopt the eddy turnover time $\tau_{\rm cs}\sim\ell_{\rm cs}/c$, while \citet[][]{NarayanPiran2012} assumed $\tau_{\rm cs}\sim \ell_{\rm cs}/\beta_{\rm rec}c$.  In our model each current sheet processes a volume $\beta_{\rm rec}c\tau_{\rm cs}\ell_{\rm cs}^2\sim \beta_{\rm rec}\ell_{\rm cs}^3$, while in their model the processed volume is $\ell_{\rm cs}^3$.

From Eq.\ \eqref{eq:tvar}, one is tempted to conclude that an arbitrary short variability timescale can be achieved in the limit $\ell_{\rm cs}\ll\ell_0$. However, from Eq.\ \eqref{eq:Ns} one sees that the number of current sheets becomes very large when $\ell_{\rm cs}\ll\ell_0$. Since each current sheet produces a pulse of the light curve, a large number of pulses may overlap, and the variability could be eventually erased. The resolution of this issue is discussed in the next section.

\subsection{Lighthouse effect}

The duty cycle of the pulsed emission can be estimated as $f_{\rm duty}\sim\min[1,\xi]$, where $\xi$ is the number of current sheets that are seen by an observer at a given time (for the compactness of notation, hereafter we refer to $\xi$ as the duty cycle). The duty cycle is crucial to assess the variability of the light curve \citep[][]{NarayanPiran2012}. When $\xi\lesssim 1$ the radiation pulses produced by different current sheets do not overlap, and the light curve shows a substantial variability. On the other hand, when $\xi\gtrsim 1$ several radiation pulses overlap, and the variability of the light curve is erased.

The duty cycle is affected by the pitch angle anisotropy of the particles energised in reconnecting current sheets. Numerical simulations of relativistic plasma turbulence show that the electron velocity is nearly aligned with the direction of the guide magnetic field, i.e.\ the maximum pitch angle $\alpha$ is small \citep[][]{ComissoSironi2019, ComissoSironi2021, Comisso+2020, SobacchiNattila2021}. Since synchrotron and IC radiation are beamed along the direction of the particle velocity, each current sheet produces a radiation beam that illuminates a solid angle $\alpha^2$. Then there is a small probability that the beam intercepts the line of sight.

The duty cycle can be estimated as follows. The probability that a current sheet is visible at a given time is $\xi/N_{\rm cs}$. This probability can be also expressed as the product of (i) the probability that the current sheet is ``active'', $\tau_{\rm cs}/\tau$, and (ii) the probability that the radiation from the current sheet is beamed along the line of sight, $\alpha^2$ (it is assumed that this probability is independent of the chosen line of sight, i.e.\ the radiation beams are distributed isotropically in the proper frame of the shell). Using Eq.\ \eqref{eq:Ns} one finds
\begin{equation}
\label{eq:xi}
\xi\sim \beta_{\rm rec}^{-1}\left(\frac{\ell_0}{\ell_{\rm cs}}\right)^2\alpha^2 \;.
\end{equation}
The pitch angle anisotropy is essential to achieve a substantial variability of the light curve. Using Eqs.\ \eqref{eq:tvar} and \eqref{eq:xi}, the condition $\xi\lesssim 1$ gives $\alpha\lesssim\alpha_1=\beta_{\rm rec}^{1/2}\eta^{-1}(\delta T/T)$. Taking $\beta_{\rm rec}\sim 0.1$ and $\eta\sim 1$, a variability timescale $\delta T/T\sim 0.1$ requires anisotropic pitch angles $\alpha\lesssim 0.03$. This level of pitch angle anisotropy is consistent with numerical simulations of relativistic plasma turbulence in the fast cooling regime \citep[][]{SobacchiNattila2021}.

Our Eq.\ \eqref{eq:xi} is similar to Eq.\ (11) of \citet[][]{NarayanPiran2012}. An important difference is that in our model the beaming of the radiation is due to the pitch angle anisotropy of the energised particles, while in their model it is due to the ultrarelativistic bulk motion of the reconnection outflow. As discussed in Section \ref{sec:comparison}, this leads to very different predictions for the counterpart of ultrafast gamma-ray flares at lower frequencies.

The duty cycle $\xi\sim N_{\rm cs}(\tau_{\rm cs}/\tau)\alpha^2$ is equal to the fraction of the solid angle that is illuminated at a given time in the proper frame. When $\xi\lesssim 1$ the isotropic equivalent of the flare luminosity inferred by an observer whose line of sight is illuminated, $L$, is larger than the average luminosity over the entire solid angle, $\langle L\rangle_{\Omega}$. The luminosity ratio is $L/\langle L\rangle_{\Omega} \sim\xi^{-1}$.

The duty cycle $\xi$ is also equal to the fraction of the shell volume filled by radiation. Since in the illuminated regions the photon number density is a factor $\xi^{-1}$ larger than average, the mean free path for photon-photon annihilation is the same as if the photons were distributed uniformly throughout the shell. Then the mean free path depends only on the total duration of the flare, and is independent of the variability timescale \citep[][]{NarayanPiran2012}.

We conclude this section with two remarks. The effective beaming of the radiation decreases if the current sheet is curved or rotates significantly. In this case the variability of the light curve tends to be erased. On the other hand, the radiation pulse produced by a current sheet may not be smooth. This would increase the variability of the light curve.

\section{Properties of the light curve}
\label{sec:LC}

\subsection{Morphology}

The number of pulses of the light curve, $N_{\rm p}$, is equal to the product of (i) the total number of current sheets, $N_{\rm cs}$, and (ii) the probability that the radiation from the current sheet is beamed along the line of sight, $\alpha^2$. Using Eq.\ \eqref{eq:Ns} and \eqref{eq:tvar} one finds
\begin{equation}
\label{eq:Np}
N_{\rm p} \sim \beta_{\rm rec}^{-1}\eta^{-1}\left(\frac{\ell_0}{\ell_{\rm cs}}\right)^3\alpha^2 \;.
\end{equation}
Using Eqs.\ \eqref{eq:tvar} and \eqref{eq:Np}, one sees that the light curve is composed of a large number of pulses, i.e.\ $N_{\rm p}\gtrsim 1$, for pitch angles $\alpha\gtrsim\alpha_2=\beta_{\rm rec}^{1/2}\eta^{-1}(\delta T/T)^{3/2}$.

The light curve has three possible morphologies, which depend on the beaming of the radiation from the current sheets.
\begin{enumerate}
\item {\it Weak beaming} $[\alpha\gtrsim\alpha_1=\beta_{\rm rec}^{1/2}\eta^{-1}(\delta T/T)]$. Since $\xi\gtrsim 1$, the light curve has one pulse of duration $T$. The variability is weak as pulses of duration $\delta T$ from different current sheets overlap.
\item {\it Intermediate beaming} $[\alpha_1\gtrsim\alpha\gtrsim\alpha_2=\beta_{\rm rec}^{1/2}\eta^{-1}(\delta T/T)^{3/2}]$. In this case one has $\xi\lesssim 1$ and $N_{\rm p}\gtrsim 1$. Then the light curve has $N_{\rm p}$ separate pulses of duration $\delta T$. This morphology has been reported in several AGN flares, including the TeV flare detected on 2006 July 28 from the BL Lac object PKS 2155--304 \citep[][]{Aharonian+2007}, and the GeV flare detected on 2015 June 16 from the Flat Spectrum Radio Quasar 3C 279 \citep[][]{Ackermann+2016}.
\item {\it Strong beaming} $[\alpha\lesssim\alpha_2]$. In this case $N_{\rm p}\sim N_{\rm cs}\alpha^2$ is the fraction of the solid angle that is illuminated during the entire flare. If the illuminated directions intercept the line of sight, the light curve has one pulse of duration $\delta T$. The TeV flares detected on 2005 June 30 and July 9 from the BL Lac object Mrk 501 \citep[][]{Albert+2007} are representative of this morphology.
\end{enumerate}

\begin{figure}{\vspace{2mm}} 
\centering
\includegraphics[width=0.46\textwidth]{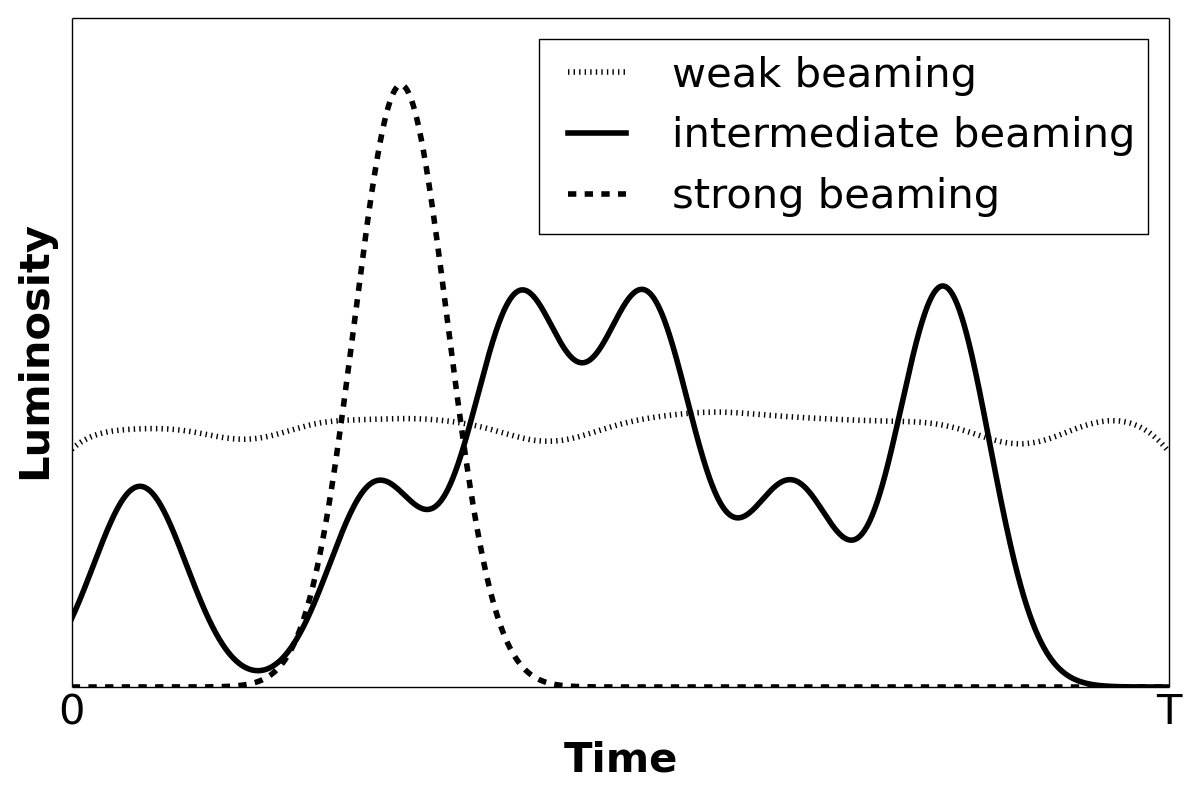}
\vspace{-3mm}
\caption{Morphology of the gamma-ray light curve for weak beaming (dotted), intermediate beaming (solid), and strong beaming (dashed). The light curve becomes more variable as the beaming of the radiation from the current sheets increases.
}
\label{fig:LC}
\end{figure}

 These three morphologies are illustrated in Figure \ref{fig:LC}. We assume that each current sheet produces a radiation pulse with a gaussian profile of full width at half maximum $\delta T=T/10$. The peak time of the pulses is drawn from a uniform probability distribution over the interval $0<t<T$. We assume $\beta_{\rm rec}=0.1$ and $\eta=1$. The light curves are obtained for pitch angles $\alpha=0.3$ (dotted), $\alpha=0.03$ (solid), and $\alpha=0.003$ (dashed), which correspond to the regimes of weak, intermediate, and strong beaming respectively. Since beaming is very sensitive to the conditions at the outer scale of the turbulent cascade \citep[][]{Comisso+2020, SobacchiNattila2021}, different flares (even produced by the same AGN) may exhibit different morphologies.

\subsection{Dependence on the observed frequency}

\subsubsection{Synchrotron vs.\ IC luminosity}

Our model predicts that the IC luminosity, $L_{\rm IC}$, is higher than the synchrotron luminosity, $L_{\rm sync}$. One has\footnote{For simplicity we do not discuss the Klein-Nishina corrections to IC scattering. The full analysis is presented in \citet[][]{SobacchiSironi2021}.}
\begin{equation}
\label{eq:LIC}
\frac{L_{\rm sync}}{L_{\rm IC}}= \left(\frac{U_{\rm B}}{U_{\rm rad}}\right)\sin^2\alpha = \left(\frac{U_{\rm B}}{U_{\rm ext}+U_{\rm sync}}\right)\sin^2\alpha \;, 
\end{equation}
where $U_{\rm B}$ is the magnetic energy density, and $U_{\rm rad}=U_{\rm ext}+U_{\rm sync}$ is the radiation energy density ($U_{\rm ext}$ is the energy density of the external photon field, and $U_{\rm sync}$ is the energy density of the sychrotron photons, which is determined self-consistently below). In the fast cooling regime, the available magnetic energy is converted into synchrotron and IC radiation within one light crossing time of the shell. Since $U_{\rm B}\sim U_{\rm sync}+U_{\rm IC}$, the energy density of the sychrotron photons can be estimated as
\begin{equation}
\label{eq:Usync}
U_{\rm sync}\sim \left(\frac{U_{\rm sync}}{U_{\rm sync}+U_{\rm IC}}\right)U_{\rm B}\sim \left(\frac{L_{\rm sync}}{L_{\rm sync}+L_{\rm IC}}\right)U_{\rm B}\;.
\end{equation}
Substituting Eq.\ \eqref{eq:Usync} into Eq.\ \eqref{eq:LIC}, one finds\footnote{The two limiting cases of Eq.\ \eqref{eq:Lratio} are obtained for different ratios $U_{\rm sync}/U_{\rm ext}$. When $U_{\rm sync}\ll U_{\rm ext}$, Eq.\ \eqref{eq:LIC} gives $L_{\rm sync}/L_{\rm IC}\sim (U_{\rm B}/U_{\rm ext})\sin^2\alpha$. When $U_{\rm sync}\gg U_{\rm ext}$, Eq.\ \eqref{eq:LIC} gives $L_{\rm sync}/L_{\rm IC}\sim (U_{\rm B}/U_{\rm sync})\sin^2\alpha$, and Eq.\ \eqref{eq:Usync} gives $U_{\rm sync}/U_{\rm B}\sim L_{\rm sync}/L_{\rm IC}$. Then one finds $L_{\rm sync}/L_{\rm IC}\sim\sin\alpha$.}
\begin{equation}
\label{eq:Lratio}
\frac{L_{\rm sync}}{L_{\rm IC}} \sim \min\left[\sin\alpha ,\; \frac{U_{\rm B}}{U_{\rm ext}}\sin^2\alpha\right]\;.
\end{equation}
For a strong pitch angle anisotropy ($\alpha\ll 1$), as required to achieve a fast variability, one has $L_{\rm sync}/L_{\rm IC}\ll 1$. Then the bolometric luminosity of the flare is dominated by the IC emission.

It is possible that the observed emission includes a non-flaring component. Since the synchrotron emission of the flaring component is suppressed due to the pitch angle anisotropy, it may be partially or completely ``buried'' under the non-flaring component. In this case fast variations of the observed synchrotron luminosity have a small amplitude.

\subsubsection{Effect of the particle cooling time}
\label{sec:cooling}

The effect of the particle cooling time on the variability of the light curve can be understood by considering the regime $\tau_{\rm cs}\lesssim\tau_{\rm cool}\lesssim\tau$. In this regime the cooling time is longer than the light crossing time of a current sheet, and shorter than the light crossing time of the shell, as appropriate in the fast cooling regime.

When $\tau_{\rm cool}\gtrsim\tau_{\rm cs}$, each current sheet produces a radiation pulse of duration $\tau_{\rm cool}$. Repeating the same arguments used in the derivation of Eqs.\ \eqref{eq:tvar} and \eqref{eq:xi}, one sees that $\delta T/T$ and $\xi$ increase by a factor $\tau_{\rm cool}/\tau_{\rm cs}$. Then the synchrotron and IC emission should be less variable at lower frequencies, as the cooling time of the radiating particles increases.

\subsection{The flares from PKS 2155--304}

The BL Lac object PKS 2155--304 entered a period of high activity in July 2006. Available estimates of the supermassive black hole mass indicate $M\sim 1-2\times 10^9M_\odot$ \citep[][]{Bettoni+2003}, which correspond to $R_{\rm g}/c=2GM/c^3 \sim 1-2\times 10^4{\rm\; s}$. An exceptional TeV flare was detected on July 28 \citep[][]{Aharonian+2007}. The variability timescale of the gamma-ray light curve was $\delta T\sim 400 {\rm\; s}$, i.e.\ a factor $c\delta T/R_{\rm g}\sim 0.02-0.04$ shorter than the light crossing time of the supermassive black hole. Unfortunately, there are no multiwavelength observations of this flare.

PKS 2155--304 flared for a second time on July 30. Simultaneous optical, X-ray, and gamma-ray observations of this flare were reported by \citet[][]{Aharonian+2009}. The variability timescale of the gamma-ray light curve was $\delta T\sim 4\times 10^3{\rm\; s}$. During the flare the bolometric luminosity was dominated by the IC emission (gamma-rays), which was brighter than the synchrotron emission (X-rays, optical) by a factor of $\sim 10$. A few hours after the flare the IC luminosity was comparable to the synchrotron luminosity. During the flare the gamma-ray luminosity varied by a factor of $\sim 20$. The X-ray luminosity varied by a factor of $\sim 2$, and the optical variations were weak ($\sim 15\%$).

These observations are consistent with our scenario. Estimating $L_{\rm sync}/L_{\rm IC}\sim 0.1$ and $\delta T/T\sim c\delta T/R_{\rm g}\sim 0.2-0.4$, Eqs.\ \eqref{eq:tvar} and \eqref{eq:Lratio} give $\eta \ell_{\rm cs}/\ell_0\sim 0.2-0.4$ and $\alpha\sim 0.1$. Assuming $\beta_{\rm rec}\sim 0.1$ and $\eta\sim 1$, Eqs.\ \eqref{eq:xi} and \eqref{eq:Np} give $\xi\sim 0.6-2$ and $N_{\rm p}\sim 1-10$, consistent with the fact that the gamma-ray and X-ray emission show a substantial variability.\footnote{For the exceptional TeV flare of July 28, one has $\eta\ell_{\rm cs}/\ell_0\sim c\delta T/R_{\rm g}\sim 0.02-0.04$. Since there are no multiwavelength observations, one cannot estimate $\alpha$ directly. The duty cycle is $\xi\sim 0.6-2\times 10^4\eta^2\alpha^2$. The condition $\xi\lesssim 1$ is satisfied for $\eta\alpha\lesssim 0.01$.} The ratio of the cooling times of the particles responsible for the optical and X-ray emission is $\tau_{\rm cool, opt}/\tau_{\rm cool, X}\sim (\nu_{\rm X}/\nu_{\rm opt})^{1/2}\sim 10^2$. Then it is not surprising that the optical emission is weakly variable.

\section{Comparison with previous work}
\label{sec:comparison}

\subsection{Models invoking relativistic bulk motions}

Models attributing the variability of the light curve to the relativistic random motion of ``blobs'' in the proper frame of the shell were developed for GRBs \citep[e.g.][]{Lyutikov2006, Lazar+2009, NarayanKumar2009} and later applied to AGN \citep[e.g.][]{Giannios2009, Nalewajko2011, NarayanPiran2012, Giannios2013}. In these models each blob dominates the emission of the whole shell when it moves towards the observer.

A common version of these models is the reconnection-driven ``jets-in-a-jet'' scenario, in which a reconnection layer is fragmented into a chain of plasmoids that are accelerated to ultrarelativistic velocities \citep[][]{Giannios2009, Petropoulou+2016, Christie+2019}. We emphasize that this scenario requires special conditions. The bulk Lorentz factor of the reconnection outflow can be estimated as $\gamma_{\rm bulk}\sim \min[\sqrt{\sigma}, \epsilon^{-1}]$, where $\sigma\gg 1$ is the plasma magnetisation, and $\epsilon$ is the ratio of the guide magnetic field and the total field \citep[][]{Lyubarsky2005, ComissoAsenjo2014}. The magnetic field across the current sheet should be nearly anti-parallel ($\epsilon\lesssim 1/\sqrt{\sigma}$) in order to achieve a large Lorentz factor $\gamma_{\rm bulk}\sim\sqrt{\sigma}$, as required in the jets-in-a-jet model. If the guide magnetic field is of the same order of the reconnecting component ($\epsilon\sim 1$), as in our turbulent scenario, one has $\gamma_{\rm bulk}\sim 1$. Current sheets with a vanishing guide field may form if the supermassive black hole accretes magnetic loops of alternating polarity \citep[][]{Lyubarsky2010, GianniosUzdensky2019}.

Simultaneous multiwavelength observations of ultrafast AGN flares can discriminate between the reconnection-driven jets-in-a-jet model and our scenario. Since the pitch angle distribution of the energised particles is assumed to be isotropic, the jets-in-a-jet model predicts that ultrafast gamma-ray flares have a bright synchrotron counterpart at lower frequencies, while our scenario predicts a weak synchrotron variability.

\subsection{Models invoking pitch angle anisotropy}

Models attributing the variability of the light curve to the pitch angle anisotropy of the radiating particles have received limited attention. Below we discuss two such models. Similarly to our scenario, these models predict that ultrafast gamma-ray flares have a faint synchrotron counterpart at lower frequencies. 

\citet[][]{Thompson2006} suggested that the fast variability of GRB light curves is due to the pitch angle anisotropy of the energised particles in a turbulent magnetised plasma. In this model the intermittency of the turbulent cascade is neglected, and consequently the dissipation rate in the emission region is uniform. Then the energised particles fill the whole shell. As we show below, in this case the light curve could hardly show a significant variability. Our argument is independent of the specific dissipation process, and therefore applies to any model where the intermittency of the turbulent cascade is neglected.

\citet[][]{Thompson2006} associated the variability of the light curve to the timescale $\tau_\alpha$ to tilt the magnetic field by an angle $\alpha$ (where $\alpha\ll 1$ is the maximum pitch angle of the energised particles). The idea is that radiation produced by particles moving along a certain field line is beamed out of the line of sight when the magnetic field is tilted. The timescale $\tau_\alpha$ can be estimated as follows. Models of Alfv\'{e}nic turbulence predict that the amplitude of magnetic field fluctuations at a certain scale $\ell$ in the inertial range is $\delta B/B\sim (\ell_\perp/\ell_0)^{1/3}\sim (\ell_\parallel/\ell_0)^{1/2}$, where $\ell_0$ is the outer scale of the cascade, and the parallel and perpendicular directions are defined with respect to the local magnetic field \citep[][]{GoldreichSridhar1995, ThompsonBlaes1998}. In order to tilt the magnetic field by an angle $\alpha$, the amplitude of the Alfv\'{e}n wave should be $\delta B/B\sim\alpha$. The corresponding wavelength is $\ell_{\alpha\parallel} \sim\alpha^2\ell_0$, which gives $\tau_\alpha\sim \ell_{\alpha\parallel}/ v_{\rm A} \sim \alpha^2\ell_0/c\ll\ell_0/c$.

The shortcoming of this model is that the duty cycle, which was not calculated by \citet[][]{Thompson2006}, is expected to be very large. The number of eddies of parallel size $\ell_{\alpha\parallel} \sim \alpha^2 \ell_0$ and perpendicular size $\ell_{\alpha\perp} \sim\alpha^3 \ell_0$ can be estimated as $N_\alpha\sim\ell_0^3/\ell_{\alpha\parallel} \ell_{\alpha\perp}^2\sim \alpha^{-8}$. Since the intermittency of the turbulent cascade is neglected, the energised particles fill the whole shell. Then in this model the duty cycle is $\xi\sim N_{\alpha} \alpha^2\sim\alpha^{-6}\gg 1$, showing that the variability of the light curve on the timescale $\tau_\alpha$ is erased.


\citet[][]{Ghisellini+2009} proposed a magnetospheric model of ultrafast AGN flares invoking pitch angle anisotropy. In this model the flares are due to IC emission by electrons that are magnetocentrifugally accelerated along the field lines. As discussed earlier, magnetospheric models cannot explain the origin of the ultrafast TeV flare from the Flat Spectrum Radio Quasar PKS 1222+21 \citep[][]{Aleksic+2011}, as the flare should be produced at a distance $\gtrsim 0.1{\rm\; pc}$ in order to prevent absorption of the gamma-rays by the UV photons from the Broad Line Region.

\section{Conclusions}
\label{sec:conclusions}

We have proposed that ultrafast AGN flares (variability timescale shorter than the light crossing time of the supermassive black hole Schwarzschild radius) are a manifestation of the intermittency of the turbulent cascade that dissipates the jet Poynting flux. In our scenario the variability timescale is equal to the light crossing time of the intermittent current sheets where the particles are energised by the reconnection electric field. Each current sheet produces a radiation pulse whose duration is shorter than the light crossing time of the whole emission region. The latter can be comparable to the light crossing time of the supermassive black hole Schwarzschild radius.

In our scenario, particles energised by reconnection have a strong pitch angle anisotropy, i.e.\ their velocity is nearly aligned with the guide magnetic field. Therefore, the radiation from each current sheet is beamed. To obtain observed variability, it is essential that the beam illuminates only a small fraction of the solid angle, so that a single current sheet dominates the emission from the whole jet. Correspondingly, if the opening angle of the beams were too broad, radiation pulses from different current sheets would overlap, and the variability of the light curve would be erased. Flares would exhibit a variety of temporal structures depending on the specific parameters of the reconnection layers.

The predictions of our model are the following:
\begin{enumerate}
\item The bolometric luminosity of ultrafast AGN flares is dominated by the IC emission. The lower energy synchrotron emission is suppressed as the velocity of the radiating particles is nearly aligned with the local magnetic field.
\item If the observed emission includes a non-flaring component, the synchrotron emission of the flaring component may be partially or completely ``buried''. Then fast variations of the synchrotron luminosity have a small amplitude.
\item The synchrotron and IC emission are less variable at lower frequencies, as the cooling time of the radiating particles exceeds the light crossing time of the current sheet.
\end{enumerate}
The first and second predictions are a distinctive signature of models invoking a strong pitch angle anisotropy of the radiating particles. Simultaneous multiwavelength observations of ultrafast AGN flares can test test these predictions, and discriminate between our scenario and the jets-in-a-jet model considered in earlier work \citep[][]{Giannios2009, Petropoulou+2016, Christie+2019}, as the latter predicts bright ultrafast synchrotron flares.

Establishing that the radiating particles have a strong pitch angle anisotropy would be important for the modelling of the Spectral Energy Distribution (SED) of relativistic jets. SED models suggest that jets from Active Galactic Nuclei are matter dominated, i.e.\ the magnetic energy density is much smaller than the energy density of the radiating electrons \citep[][]{TavecchioGhisellini2016}. Interestingly, recent attempts to model the SED of jets from Tidal Disruption Events have reached a similar conclusion \citep[][]{Pasham+2023}. These result are in tension with the theoretical prediction that relativistic jets are magnetically dominated objects \citep[][]{BlandfordZnajek1977, Tchekhovskoy2011}. The tension is alleviated if the radiating particles have small pitch angles \citep[][]{SobacchiLyubarsky2019, SobacchiSironi2021}. Since the synchrotron frequency and power depend on the component of the magnetic field perpendicular to the particle velocity, $B\sin\alpha$, the strength of the jet magnetic field inferred from the SED is very sensitive to the pitch angle anisotropy. The magnetic energy density could be underestimated by a factor $\sin^2\alpha\ll 1$ by SED models making the standard assumption that the pitch angle distribution is isotropic.

\section*{Acknowledgements}

We acknowledge fruitful discussions with Daniel Groselj, Yuri Lyubarsky, and Fabrizio Tavecchio. TP and ES are supported by an advanced ERC grant MultiJets and ISF grant 2126/22. LC is supported by NASA ATP 80NSSC22K0667.

\def\aap{A\&A}\def\aj{AJ}\def\apj{ApJ}\def\apjl{ApJ}\def\mnras{MNRAS}
\def\prl{Phys. Rev. Lett.}
\def\araa{ARA\&A}\def\physrep{PhR}\def\sovast{Sov. Astron.}\def\nar{NewAR}\def\pasa{PASA}
\def\aapr{Astronomy \& Astrophysics Review}\def\apjs{ApJS}\def\nat{Nature}\def\na{New Astron.}
\def\prd{Phys. Rev. D}\def\pre{Phys. Rev. E}\def\pasp{PASP}\def\apss{ApSS}
\bibliographystyle{aasjournal}
\bibliography{2d}


\end{document}